\documentstyle[11pt,newpasp,twoside]{article}
\def\edcomment#1{\iffalse\marginpar{\raggedright\sl#1\/}\else\relax\fi}
\reversemarginpar
\def \yskip{\penalty-50\vskip3pt plus 3pt minus 2pt}
\def \reference{\par \yskip \noindent \hangindent .4in \hangafter 1}
\def \abc#1#2#3#4 {\reference#1, {\sl#2}, {\bf#3}, #4}
\def \blank {\lower 5pt\hbox to 0.75in{\hrulefill}}

\def \cm{\rm{cm}}

\def \s{\rm{s}}
\def \km{\rm{km}}
\def \lae{\mathrel{<\kern-1.0em\lower0.9ex\hbox{$\sim$}}}
\def \gae{\mathrel{>\kern-1.0em\lower0.9ex\hbox{$\sim$}}}
\def \yskip{\penalty-50\vskip3pt plus 3pt minus 2pt}
\def \reference{\par \yskip \noindent \hangindent .4in \hangafter 1}
\def \abc#1#2#3#4 {\reference#1, {\sl#2}, {\bf#3}, #4}
\def \blank {\lower 5pt\hbox to 0.75in{\hrulefill}}

\def \cm{\rm{cm}}

\def \s{~\rm{s}}
\def \km{~\rm{km}}
\def \pc{\rm{pc}}
\def \lae{\mathrel{<\kern-1.0em\lower0.9ex\hbox{$\sim$}}}
\def \gae{\mathrel{>\kern-1.0em\lower0.9ex\hbox{$\sim$}}}

\begin{document}
\title{Theory of the Interaction of Planetary Nebulae with the Interstellar Medium}
 \author{Ruth Dgani}
\affil{Department of Astronomy, The University of Texas at Austin, Austin TX 78712 }

\begin{abstract}
The theory of the interaction of planetary nebulae with the interstellar 
medium is important for the interpretation of nebular morphologies that 
deviate from point symmetry. It can be used to probe the interstellar 
medium and its magnetic field. We emphasize in this review the role of
hydrodynamical instabilities in the interaction. 
\end{abstract}

\section{Introduction}
The interaction of planetary nebulae (PNs) with the interstellar medium (ISM)
was first discussed by Gurzadyan (1969). Smith (1976) used the thin shell model
to analyze the evolution of a spherically symmetric expanding PN shell
moving through the ISM (see also Isaacman 1979). 
On the observational side, only a few PNs showed interaction with the ISM 
(hereafter refereed to as interacting PNs) until the last decade.
With better observational imaging instruments (CCD cameras) the field
interacting PNs became more popular 
(e.g., Borkowski, Tsvetanov, \& Harrington 1993; Jacoby \& Van de Steene 1995;
Tweedy, Martos \& Noriega-Crespo 1995; Hollis {\it et al.} 1996;
Zucker \& Soker 1997). Borkowski, Sarazin \& Soker (1990;hereafter  BSS) discuss
the potential in using interacting PNs to probe the ISM, and list many 
interacting and possibly interacting PNs.
This list was extended in the review article by Tweedy (1995)
and in the atlas of interacting PNs (Tweedy \& Kwitter 1996). Tweedy 
with several collaborators has been leading the observational research
in the field (e.g., Tweedy \& Kwitter 1994a,b; Tweedy \& Napiwotzki 1994).
Other studies are in progress, and preliminary 
results were presented in the conference (e.g. Kerber {\it et al.} 2000;
Dopita {\it et al.} 2000 ).

On the theoretical side, Soker, Borkowski \& Sarazin (1991;hereafter
SBS) performed numerical simulations, both in the isothermal and adiabatic 
limits.
Their results validate the thin-shell approximation of Smith (1976).
In the adiabatic case they found that the Rayliegh-Taylor (RT)
instability plays a crucial role in the evolution.
Recent numerical calculations reported in this conference 
(Dopita {\it et al.} 2000; Villaver {\it et al.} 2000) confirmed these 
results. In their observational analysis of the interacting PN IC 4593,
Zucker \& Soker (1993) suggested that the interaction process oscillates
between the adiabatic and isothermal cases.
An analytical study of this process, the so-called radiative shock
overstability, was performed by Dgani \& Soker (1994) for finite-sized
objects (e.g., spherical nebulae).
The radiative overstability was found to occur only in large
PNs, with radii of $R \gae 10^{19} n_0^{-1} \cm$, where $n_0$ is the
total number density of the ISM (in $\cm^{-3}$)
The instability occurs only for large PNs because  the transverse
flow, here it is the flow around the nebula, stabilizes the radiative
overstability (Dgani \& Soker 1994; see also
Stevens, Blondin \& Pollock 1992, for a numerical study of a different
situation). 
For an earlier review of the theory see Dgani(1995).
Several recent theoretical studies were conducted since that review.  
Soker \& Dgani (1997) studied analytically the interaction
of PNs with a magnetized ISM. Dgani \& Soker (1998) and Dgani(1998)  discussed
the role of instabilities in the interaction. 
New numerical studies are underway (Villaver {\it et al.} 2000; 
Dopita {\it et al.} 2000), preliminary results presented in the conference
show that instabilities play an important role in the interaction.

\section{Preliminary Considerations}

\subsection{\it Why is the interaction important?}
The interaction of PNs with the ISM causes deviation form axisymmetry 
(or point symmetry). Several other processes
that cause deviation from axisymmetry involving binaries were discussed by 
Soker and collaborators (e.g. Soker 1994; Soker 1996; Soker et al 1998).

It is important to understand as accurately as possible the ISM effects on
the morphology and to distinguish them from other processes.

Interacting PNs are also a tool for studying the ISM. This role was
emphasized in previous reviews (Tweedy 1995; Dgani 1995).

\subsection{\it Detecting the interaction}
The interaction can be divided into at least three phases: free expansion,
deceleration and stripping off the central star. Initially, PNs expand freely
because their densities are of orders of magnitude higher than the densities of 
their surrounding medium. Eventually, though, every nebula will reach a stage
in which the pressure inside the nebula is of the same order as the pressure
outside the nebula.
 A strong enough pressure wave will then propagate into the nebula and 
decelerate it.
If the nebula moves supersonically, the dominant ISM pressure is the ram
pressure.  The critical density $n_{crit}$, at which the ISM and the nebula 
reach the same pressure is (BSS, Eq. 1):
 \begin{eqnarray}
  n_{crit} =\left({v_*+v_{e}\over c}\right)^2 n_{0} \;, 
\end{eqnarray}
where $c$ is the isothermal sound speed in the nebula,
$c\sim $ 10 km s$^{-1}$
in most cases, $v_*$ is the velocity of the central star,
$v_{e}$ is the expansion velocity of the nebula
and $n_0$ is the ISM density.
The most difficult parameter to observe directly  in the above
formula is the ISM density, $n_0$.
The nebula at this early stage of the interaction is
not distorted, but its leading edge is brighter.  Measuring the
value of the density at the leading edge, $n_{crit}$, and using the  
above formula yields an estimate for $n_0$.

\noindent
If we take the average value of $v_{e}=$ 20 km s$^{-1}$ (Weinberger 1989)
and $v_*= $ 60 km s$^{-1}$ (Pottash 1984; BSS), then $n_{crit}=60 n_0$.
As the ISM ram pressure depends on the velocity squared,
the interaction is much stronger for fast moving planetaries.
For example, if $v_*=$ 150 km s$^{-1}$  then the critical
density is about $300$ times that of the ISM.

\noindent
The electron number density of extended planetaries is about 100 cm$^{-3}$
(Kaler 1983) for the low surface brightness PNs from the NGC catalog,
and 10 cm$^{-3}$ for the fainter PNs from the Abell (1966) catalog.
The very large nebula S-216 has $n_e=$5 cm$^{-3}$.
In principle, therefore, fast moving planetary nebulae can be detected in
very low density environments $n_{0}\simeq 0.01$.

\section{Numerical Models}
\noindent
The first realistic models of moving PNs were calculated numerically by
SBS.  They performed hydrodynamic simulations of the PN-ISM interaction
with the particle in cell (PIC) method in cylindrical symmetry. They
calculated two types of models of thick nebular shells moving through the ISM:
an adiabatic model which applies to fast nebulae moving in the
galactic halo and an isothermal model relevant to average nebulae
moving in the galactic plane.
\noindent

\subsection{\it Cooling in the ISM shock}
A nebula moving supersonically in the ISM creates a strong bow
shock before it. The cooling time of the shocked
ISM gas depends on the ionization state of the gas before
the shock. The following analytical formula for the cooling
time is based on numerical calculations of a steady shock
with ionization precursor. It is accurate within a factor
of 2 for $v_s>$60 km s$^{-1}$  Mckee (1987):
\begin{eqnarray}
t_{cool}=100 \left({v_s\over 50\; km/s}\right)^3 n_{0}^{-1}\; yr.
\end{eqnarray}
In steady state shocks with velocities of about 50 km s$^{-1}$,
the cooling drops by a factor of 10 when the material is preionized
(Raymond 1979). However, the cooling time is still shorter
than the flow time for an average nebula moving in the galactic plane.
Only when the relative velocity is high and the density is low,
will an adiabatic flow appear.

\subsection{\it The adiabatic model}
SBS  choose the parameters for the adiabatic model 
so that the cooling time is larger than the flow time.
An instability appears near the axis  while the shell is  decelerating which
is interpreted as a Rayleigh-Taylor instability in the decelerated shell.
When the shocked ISM cannot cool,  the deceleration of the dense nebular shell
by the dilute shocked ISM is Rayleigh-Taylor unstable.
The main morphological features of the adiabatic flow are the ISM shock
front, the decelerated PN shell and the RT instability.
The last manifests itself as a "bump" and a hole in the nebula
along the upstream symmetry axis.
The numerical scheme used by SBS, though, being of 2D cylindrically
symmetric nature and limited spatial resolution, forces the fastest
growing mode to be of large wavelength and near the symmetry axis.
Similar instability modes are obtained in numerical simulations of
other types of dense bodies moving through an ambient media
(Brighenti \& D'Ercole 1995 for Wolf-Rayet ring nebulae, and
Jones, Hyesung \& Tregillis 1994 for dense gas clouds).
Brighenti \& D'Ercole (1995) present a test run where the numerical grid
is Cartesian rather than cylindrical. They find similar fragmentation of 
the flow, but not along the symmetry axis.

\subsection{\it The isothermal model}
In the isothermal model, efficient cooling is assumed.
The cooling is calculated in the following way:
for every cell in which the temperature is more than
10$^4$K the temperature is set equal to 10$^4$K.
The model parameters are typical for an average nebula moving 
in the galactic plane.

\noindent
In this case the shocked region is thin and "arms" of
denser ISM material follow the nebula.
The Rayleigh-Taylor instability does not appear in this case.
This is because the shocked ISM is cooled and its density is of the
same order as the nebular density.
The model agrees with the thin shell  approximation of Smith. 

\subsection{\it Recent numerical simulations.}
Numerical simulations show that several types
of instabilities which develop on the interface of spherically expanding
shells (or winds) moving with respect to the ISM can fragment the shells
(e.g. Brighenti \& D'Ercole 1995 ).
The detailed 2D numerical simulations of Brighenti \& D'Ercole (1995)
nicely show how RT and KH instabilities on the interface of the wind and
the ISM develop, and allow the ISM to penetrate into the wind-bubble.

New numerical studies of interacting PNs are underway 
(Villaver {\it et al.} 2000; Dopita {\it et al.} 2000). 
Preliminary results presented in the conference show that the RT instabilities 
can fragment the nebular shell and allow the ISM to penetrate to the inner
parts of the nebulae in some cases.  

\section{The Role of the ISM Magnetic Field in Interacting PNs}
The important role of the ISM magnetic field on shaping interacting PNs 
was  first mentioned in the  observational paper  of Tweedy {\it et al}  (1995)
about the nebula Sh 216. The morphology of the nebula  is dominated by a bow shock
followed by several parallel elongated filaments. 
Tweedy {\it et al.} (1995) noted that the magnetic pressure  of the ISM
is negligible compared to its ram pressure for the assumed central star velocity
($v_*\sim 20 $ km s$^{-1}$). They have argued that the central star velocity must 
be very low i.e. $v_*\sim 5 $ km s$^{-1}$ in order to explain the magnetic 
shaping of the nebula. Dgani \& Soker (1998; table 1) noted that too many 
interacting PNs (a dozen) show signs of magnetic shaping i.e. parallel 
elongated filaments.  It is very unlikely that all of them have such abnormally  
low central star velocity. 

\subsection{\it General features}
Soker \& Dgani (1997) conduct a theoretical study of the processes involved when 
the ISM magnetic field is important in the interaction.
In the case where the ISM is fully ionized, we define four characterizing
velocities of the interaction process:
the adiabatic sound speed $v_s=(\gamma kT/\mu m_H)$ and the Alfven
velocity $v_A = B_0/(4 \pi \rho_0)^{1/2}$ of the ISM, the expansion
velocity of the nebula $v_e$, and the relative velocity of the PN
central star and the ISM $v_\ast$.
The interesting cases, with the magnetic field lines being at a large angle
to the relative velocity direction, are: \\ 

\begin{enumerate}
\item $v_\ast \gg v_A \sim v_s \sim v_e$, and a rapid cooling behind the
shock wave.
Both the thermal and magnetic pressure increase substantially behind
a strong shock. If radiative cooling is rapid, however, the magnetic pressure
will eventually substantially exceed the thermal pressure, leading to
several strong MHD instabilities around the nebula, and probably to
magnetic field reconnection behind the nebula. 

\item $v_\ast \gg v_A \sim v_s \sim v_e$, and negligible cooling behind the
shock. The thermal pressure, which grows more than the magnetic pressure in a
strong shock, will dominate behind the shock.
 Magnetic field reconnection is not likely to occur behind the nebula.
This domain characterizes the interaction of the solar wind with
the atmospheres of Venus and Mars (e.g., Phillips \& McComas 1991). 
\end{enumerate}

In the ISM it is likely that $v_s \sim v_A$, and
$v_e \simeq 10 \km \s^{-1}$ will in most cases not exceed the ISM sound speed.
The central star velocity $v_\ast \gg v_s$ for most cases.
Case (1) is the magnetic parallel of the isothermal model (section 3.1)
which is relevant for planetary nebulae moving in the galactic plane.
Case (2) is relevant for galactic halo PNs..

\subsection {\it The magnetic Rayliegh-Taylor instability for galactic plane PNs}
The isothermal case (case 1) is the more typical for the majority of PNs,
moving moderately fast ($v_\ast\sim 60\;km/s$) close to the galactic plain,
where the ISM density is relatively high.
Soker {\it et al.} (1991) find in their numerical simulations that the
isothermal model shows no sign of the RT instability.
Soker \& Dgani (1997) show   the ISM magnetic field, which was not incorporated in
the numerical simulations of Soker {\it et al.} (1991), can lead to an
interesting manifestation of the RT instability in the isothermal case  (case 1).
We assume that there is a magnetic field of intensity $B_s$ in the shocked
ISM, but not in the nebula, and that its field lines are parallel to the
boundary separating the ISM and nebular material.
The acceleration $g$ (deceleration in the frame of the central star)
of the nebular front (pointing downstream) is in the opposite
direction to the density gradient.
In the linear regime the growth rate of the RT instability, $\sigma$, with a
wavenumber $k = 2 \pi / \lambda$, is given by  (e.g., Priest 1984; \S 7.5.2)
\begin{eqnarray}
\sigma^2 = - g k {
{\rho_n-\rho_I}
\over
{\rho_n+\rho_I}}
+
{{B_s^2}\over{4 \pi ( \rho_n+\rho_I )}}
k^2 \cos^2 \theta,
\end{eqnarray}
 where $\rho_I$ and $\rho_n$ are the ISM and nebular density, respectively,
$\theta$ is the angle between the magnetic field lines and the
wave-vector, and RT instability occurs when $\sigma^2 < 0$.
The deceleration of the leading shell
{ { by the ISM ram pressure}} is given by
\begin{eqnarray}
g\sim \pi R^2 \rho_0 v_\ast^2/M_F,
\end{eqnarray}
where R is the nebular radius, and $M_F$ is the mass being decelerated.
Taking $R=0.5 \pc$, $v_\ast=60$ km s$^{-1}$, $\rho_0= 10^{-25}$ g cm$^{-3}$ and
$M_F=0.05 M_\odot$
we obtain $g \sim 3\times 10^{-7}$ cm  s$^{-2}$.

We assume that after it is shocked the nebular material reaches
$10^4\;K$, and that its thermal pressure $\rho_n v_{sn}^2$ is approximately
equal to the ram pressure of the ISM,  $\rho_0 v_\ast^2$.
 Here $\rho_{n}$ is the shocked nebular density,
$v_{sn} \simeq 10 \km \s^{-1}$ is the isothermal sound speed in the
shocked cooled nebula, and we use the assumption $v_\ast \gg v_e$.
 The post-shock nebular density is therefore
\begin{eqnarray}
\rho_{n} = \rho_0 v_\ast^2/ v_{sn}^2.
\end{eqnarray}

RT instability occurs when the density of the shocked cooled ISM 
$\rho_s$ is smaller
than the post-shock nebular density. 
This can happened for a magnetic shock when the magnetic pressure
 provides the support for the shocked ISM region.For a fast enough growth of
RT instability we require $\rho_s/\rho_n < 1/3$.  
Assuming equipartition  in the 
undisturbed ISM, we find the condition for a strong enough RT.
Namely, we  find that for a typical PN velocity $v_\ast=60$  km s$^{-1}$
 through the ISM the pre-shocked ISM
magnetic field line should be within $45^\circ$ of the shock front direction.
 Therefore, in most typical cases, i.e., equipartition and
$v_\ast \sim 60 \km \s^{-1}$, fast growing RT instability modes with long
wavelengths will develop.

The magnetic field has a destabilizing effect only for modes
having wave numbers close to being perpendicular to the field lines
(and in the plane separating the ISM and the nebula).
Instability modes having wave numbers along the field lines are
suppressed by the magnetic tension. 
This effect may create elongated structures in the direction of the
magnetic field lines.

\subsection{\it The morphology of interacting PNs and their galactic latitude}
Dgani \& Soker (1998) explore the observational consequences of the development
of RT in galactic plane PNs moving in a magnetized ISM.
In order to define the galactic plane and galactic halo population we
use the expression for the cooling time after the ISM shock (eq. 2).
The cooling will be effective when the flow time of shocked ISM around
the nebula $t_{flow}\sim  R/v_*$ is longer than the cooling time.
Calling the ratio of the two times $\eta$, we obtain:
\begin{eqnarray}
\eta=t_{cool}/t_{flow} \sim 0.1
\left( {{v_\ast}\over{50 \km \s^{-1}}} \right)^4
\left( {{R}\over{10^{18} \cm}} \right) ^{-1}
\left( {{n_0}\over{0.1 \cm^{-3}}} \right) ^{-1}
\end{eqnarray}
Halo PNs are expected to have a speed of $\gae 100 \km \s^{-1}$ through
the ISM, while those in the plane move much slower.
In addition, close to the galactic plane the density is higher
than in the halo.
We therefore expect that halo PNs will have $\eta > 1$ and therefore
the flow will be adiabatic, while for disk PNs
$\eta < 1$ and the flow will be isothermal.
Several high quality observations,
most notably given in the Atlas published by Tweedy \& Kwitter (1996)
provide us with a sample of 34  deep  images of interacting PNs.
In table 1 of Dgani \& Soker (1998),
we give a list of the objects and their references.
Some nebulae
have several parallel stripes.
We define a nebula as striped only
if it has at least two parallel long filaments.
All of the 12 striped nebulae  are close to the galactic plane ($|z|<$250 pc).
The fact that  the striped nebulae are confined to the galactic plane
is compatible with the the  prediction
that RT will be effective for
galactic plane PNs moving in the warm medium, because of
magnetic effects, but only for modes perpendicular
to the direction of the magnetic field.
The result is elongated structures or  "RT rolls".
For galactic  halo interacting PNs the cooling is
inefficient and magnetic pressure is negligible.
RT will be effective but it will form fingers or blobs and not stripes.

\subsection{\it Measuring the ISM magnetic field with interacting PNs}
The group of  12 nebulae that show stripes  is not homogeneous; some have
thick loose  stripes, some have thin densely packed  stripes.
In Dgani (1998)  the properties
of the striped nebulae  and their application to the
study of the ISM are explored.
In particular, a simple model for the shocked ISM region is used to
derive a relation between the distance between adjacent stripes
and the strength of the magnetic field of the ISM.
If $\Delta z$  is the average spacing of the stripes
in units of the radius of the nebula  then
(see Dgani 1997, eq. 4):
\begin{eqnarray}
\Delta z\sim v_{A0} \cos{\alpha_0}/v_*,
\end{eqnarray}
where $v_{A0}$ is the Alfven speed in the ISM,  $\alpha_0$ is the angle
between the pre-shock magnetic field and the shock front, and
$v_*$ is the velocity of the central star.
According to the formula nebulae with densely packed stripes either
move faster or move in a weaker magnetic field than the ones with
thicker looser stripes.
Applying this simple formula to several striped nebulae, Dgani shows that
information about the strength of the ISM magnetic field in the local
neighborhood of these nebulae can be extracted by  accurate
observations of the members of this group. \\

{\bf Acknowledgement:}
{\it It is a pleasure to  thank Noam Soker for
introducing me to the
fascinating subject of interacting nebulae and
 for a very  pleasant collaboration.}

\end{document}